\documentclass[twocolumn,twoside,prd,floatfix,letterpaper]{revtex4}

\usepackage{ifpdf}
\ifpdf 
  \usepackage[pdftex]{graphicx} 
\else 
  \usepackage{graphicx} 
\fi
\usepackage{amsmath}
\usepackage{amssymb}
\usepackage{fancyhdr}
\usepackage{color}
\usepackage{rotating}
\usepackage[colorlinks,hyperindex]{hyperref}

\begin{document} 

\title{Blunt Honesty, Incentives, and Knowledge Exchange}
\author{Bruce Knuteson}
\email{knuteson@kn-x.com}

\begin{abstract}
We propose a simple mechanism to facilitate the buying and selling of useful, bluntly honest information.  The for-profit, arm's length knowledge exchange this mechanism enables may dramatically increase the pace of scientific progress.
\end{abstract}

\maketitle
\tableofcontents

\section{Motivation}
\label{sec:Motivation}

The past two decades have witnessed an explosion of valuable information~\footnote{The words ``information'' and ``knowledge'' are used interchangeably in this article.} easily accessible to billions of people.  The breadth and depth of quality information available for free, at your fingertips with just a few keystrokes, is breathtakingly world changing.  Extraordinary examples of individuals and organizations freely sharing information to propel advances in science, industry, design, and human well-being are all around.  Information hoarding by companies and elites is rapidly disappearing in an era of increasing and unprecedented transparency.  Information, which has always wanted to be free, is.  Content providers, forced into a greater reliance on advertising revenue, maintain a clear line between advertisements and the objective integrity of journalistic content.  This remarkable transition in the way information is produced and disseminated is one of the most profound in recent human history, and one with powerful implications for our future.  After thousands of years of backbreaking labor, we now live and work in an Information Economy.

Moreover, we live in an Information Economy in which information is free.

Against this backdrop, this article's purpose is limited and superficially regressive.  We point out a small subset of information exchange in which both parties may benefit from the exchange not being free.  The scenario in Section~\ref{sec:QandA} illuminates this limited subset of information exchange while simultaneously motivating a protocol designed to facilitate such exchanges.  Section~\ref{sec:Summary} summarizes.

\section{Q\&A}
\label{sec:QandA}

Scientist A, who recently discovered an interesting, not-yet-published property of a particular molecule, hears through the grapevine that her discovery might be valuable to Scientist Q, who has been working for years to develop a cure for a rare disease~\footnote{A fanciful example involving two entrepreneurial scientists is convenient for the purpose of this article.  Real world usage is not limited to science.  The information exchange protocol described here should be useful to law enforcement (bridging the chasm between free tips and tips worth large bounties), governments, regulators, businesses, and individuals, in many ways.}.  The two scientists Q and A work in different fields, do not know each other, and have no common contacts.

A wants to sell her knowledge to Q for \$100K.  Q will happily pay \$100K for any information likely to help her complete her ambitious research program.  The obvious transaction, unfortunately, is problematic.  Before Q agrees to pay \$100K for A's information, Q wants some guarantee of the usefulness and accuracy of A's information.  The only way A can convince Q the information is worth paying for is by telling Q the information.  If A tells Q the information before the transaction is agreed upon, Q has no reason to agree to the transaction.  Both Q and A understand all of this, so A never approaches Q with her discovery, Q gets hit by a bus on a sunny afternoon two months later, the cure nearly within Q's grasp goes undiscovered for another fifteen years, and some ten thousand people and their families worldwide, including someone you love, suffer unnecessarily during the intervening decade and a half.

Our society's existing procedures for dealing with cases like this, often involving legal contracts, can be expensive to construct, monitor, and enforce.  Perhaps we can devise something better.

To assure Q of the usefulness of A's information, Q and A can agree that Q is interested in an answer to the question ``What molecule will enable the following reaction?''  Important details are clarified in a couple of accompanying paragraphs.  Q and A agree that a valid answer must be ``a molecule, fully specified in its structure.''  Agreeing on the category of valid answers is very important to Q, who has no desire to pay \$100K to be told the answer is ``a molecule similar in character to one that can be found in the excrement of some insects'' -- which, while perhaps true, is of little use to Q.  Q and A agree on a third party, X, who will decide whether A's answer falls within the agreed upon category of allowed possible answers.

Q will only agree to the transaction if she considers A's incentives to be well aligned with providing a bluntly honest answer~\footnote{The difference between ``honest'' and ``bluntly honest'' is not worth putting too fine a point on, but it is perhaps worth a quick note.  If a friend comes to you with a silly business idea and you express skeptical support, you are being honest.  You are being bluntly honest if, after some due diligence, you tell him you think the probability he closes up shop within two years is 98\%, and you are willing to stake some money on that.  If a colleague or employer asks you a question and you provide a direct answer, glossing over some complications that can be viewed as tangential to the specific question you have been asked, you are being honest.  If you also note, in detail, some related, uncomfortable fact you feel is important, knowing that doing so may jeopardize your relationship, you are bluntly honest.}.  A must have some skin in the game.  A must be rewarded if correct, and penalized if incorrect~\footnote{If A has the possibility of reward but no possibility of penalty (roughly the incentives applying to those currently managing your money), A has a clear economic incentive to enter into whatever transactions she can, whether or not she knows anything.}.

The social and reputational incentives guiding the actions of individuals can be difficult to understand, and harder still to engineer.  Anonymizing Q and A conveniently mitigates these incentives.  Nobody other than X, who brokers the transaction between Q and A, needs to know the identities of Q and A (or their agents).  This anonymization is liberating, allowing Q to ask questions without worrying about appearing ignorant, and allowing A to provide a bluntly honest answer without having to worry about delivering insult or saving face.

With social and reputational incentives thus sidelined, money is the natural carrot and stick.  A must lose money if she is wrong and make money if she is right.  For this to work, Q and A must agree up front on a date D by which the correctness of A's answer will be determined, and X must hold money from Q and A in escrow.

X, who is generally competent but not omniscient, lacks the resources necessary to determine whether A's answer is correct.  X must outsource this task, demanding accompanying evidence sufficient to convince X beyond reasonable doubt.  X cannot outsource this task to A, who has a clear incentive to claim her answer is correct.  However, {\em provided the question is one for which Q will be able to verify the accuracy of any answer received, X can reasonably outsource this task to Q}.  At the start of the transaction, X thus demands from Q an additional deposit that will be returned to Q if, by the agreed upon date D, Q informs X of the accuracy (or inaccuracy) of any answer received, providing enough evidence to convince X beyond reasonable doubt.

With Q and X thus responsible for determining the correctness of A's answer, A will rationally mistrust the game and decline to play if Q or X can make money from A's loss.  X must therefore make the same amount whether A's answer turns out to be right or wrong, and Q must pay the same amount whether A's answer turns out to be right or wrong.  Q, who consults doctors, lawyers, accountants, mechanics, and other experts when needed, is familiar with the concept of paying the same amount whether A's answer turns out to be right or wrong.

With X responsible for determining whether Q gets her deposit back, Q will rationally mistrust the game and decline to play if X can make more money by denying the sufficiency of the evidence Q provides.  X must therefore agree up front to take a fixed fee.  The amount X makes cannot depend on whether X approves or denies Q's claim, nor on whether Q claims A's answer is right or wrong.

In some cases, satisfying the above constraints will result in money that cannot be given to Q, A, or X.  They agree up front to donate any such money to charity~\footnote{``Charity'' can be anyone other than Q, A, and X.}.

Putting some simple numbers to this story may help give it substance.  Q gives X \$100, plus an additional \$100 deposit.  A gives X \$50, together with A's answer to Q's question.  X passes A's answer along to Q, respecting the anonymity of both Q and A.  X pockets \$50 for brokering the transaction~\footnote{Simple numbers are chosen for this exposition.  The equality of Q's deposit and the amount Q is willing to pay is not necessary, but it is of the right scale:  the deposit must be large enough to get Q to inform X of the accuracy of A's answer, while not being so large it unduly deters Q from entering into the transaction in the first place.  Of the \$100 Q is willing to pay, the split of \$50 to A and \$50 to X is one of similar expositional convenience.}.  There are now three possibilities.
\begin{itemize}
\item On or before date D, Q informs X that A's answer is correct, providing sufficient evidence to convince X beyond reasonable doubt.  X returns Q's \$100 deposit.  X returns A's \$50.  X pays A an additional \$50.
\item On or before date D, Q informs X that A's answer is wrong, providing sufficient evidence to convince X beyond reasonable doubt.  X returns Q's \$100 deposit.  X sends \$100 to charity.
\item Date D passes without Q getting back to X, or without Q providing sufficient justification to convince X beyond reasonable doubt.  X returns A's \$50.  X sends \$150 to charity.
\end{itemize}

A's incentives are straightforward.  A makes \$50 if she is right.  A loses \$50 if she is wrong.  A is incentivized to enter into the transaction only if she is pretty sure she is right.

X's incentives are similarly straightforward.  X gets \$50 whether A is right or wrong, and whether or not Q submits a claim with evidence sufficient for X to approve.  X, whose business depends on brokering future information exchanges, is incentivized to objectively weigh Q's evidence, and to do whatever is necessary to ensure transactions proceed smoothly and efficiently.

Q's incentives are also straightforward.  Q pays \$100 for her answer whether A is right or wrong.  Q has no economic incentive to claim A's answer is wrong.  Q has an economic incentive (her \$100 deposit) to inform X of the accuracy of the answer she received, but no economic incentive~\footnote{Here and elsewhere, ``economic incentive'' refers to a monetary incentive within the context of the paid information exchange protocol developed in this article.  Very real economic incentives external to this protocol may lead A to provide an intentionally incorrect answer, and may lead Q to lie about the accuracy of the answer(s) received.  The information exchange protocol developed in this article attempts to mitigate these external economic incentives.  The protocol does not eliminate them.} to lie about the accuracy of that answer.  If Q is unable to verify the accuracy of the answer received, Q does have an economic incentive (her \$100 deposit) to make something up and provide fake evidence to support it, but this is balanced by X, who has the power to approve or deny Q's claim, and whose continuing business interest relies on her consistently and objectively assessing the quality of the evidence Q provides.

An even more straightforward case is obtained if Q can specify up front how the accuracy of A's answer is to be determined.  Suppose Q lays out a simple procedure, easily followed by X, for determining the accuracy of any answer Q receives.  Further suppose Q, X, and A all agree, up front, that X will follow this procedure to determine the accuracy of whatever answer A provides.  X, who no longer needs Q to eventually say whether A's answer is accurate, can go ahead and return Q's \$100 deposit.  Moreover, with Q playing no further role in the adjudication process, it is fine to give Q her money back if A turns out to be wrong.  That is, {\em{if Q can specify up front, to the satisfaction of A and X, exactly how X will determine the accuracy of A's answer}}, then X makes \$50 and A stands to make or lose \$50, as above, but {\em{there is no need for the charity, and Q gets all of her money back if A turns out to be wrong}}.

This simple protocol thus interlocks the self-interested individuals Q, A, and X in a manner facilitating the transfer of useful, bluntly honest information from A to Q.  In this particular scenario, Q tests the answer provided by A, A turns out to be correct, money changes hands as prescribed, a cure is obtained, and the quality of many lives dramatically improves.

\section{Summary}
\label{sec:Summary}

This article has identified a subset of information exchange in which both parties may benefit from the exchange not being free.  The subset identified~\footnote{Many types of information transfer obviously lie outside the scope of this protocol.} are exchanges that can be cast as simple questions with an easily agreed upon set of possible answers for which Q will eventually be able to inform X of the accuracy of any answer received.  Clear specification of the category of allowed possible answers before the transaction protects Q against receiving an answer that is true but unhelpful.  Restricting questions to those for which Q will eventually be able to verify the accuracy of any answer received allows the use of an information exchange protocol, described in Section~\ref{sec:QandA}, designed to incentivize the transfer of useful, bluntly honest information from A to Q, brokered by X.

The protocol attempts to align, with robustness suitable for practical use, the incentives of very human buyers and sellers of information.  The resulting alignment, although imperfect, may represent a significant improvement on the incentives created by other communication channels.  The protocol does not guarantee correct answers, but it does make it costly for A to provide an incorrect answer.  The protocol relies on a central adjudicator X, but limits her role to approving the evidence backing the claim eventually provided by Q, and provides a strong and clear economic motivation for X to consistently and objectively weigh the evidence put in front of her.  Q is motivated to ask a clear question that will elicit an answer helpful to her.  A is motivated to carefully consider the question and provide an accurate answer within the specified category of valid possible answers.  The protocol itself is little more than a few rules, of roughly the same level of complexity as a typical board game, that create and enforce a simple incentive structure~\footnote{For most board games, reading the directions is insufficient.  You need to play a few times.}.

We hope the paid knowledge exchange developed in this article may supplement, in some small but useful way, the vast trove of information that has already been unleashed by our remarkably vibrant, mostly honest, and ostensibly wondrously robust Free Information Economy.

\acknowledgments

Kn-X~\footnote{\url{http://Kn-X.com}}, a for-profit knowledge exchange, has been a valuable platform for exploring, refining, and testing an extended version of this protocol~\footnote{Patent pending.}.  This work has benefitted from comments, suggestions, and criticisms levied by a diverse set of friends, colleagues, and critics~\footnote{Georgios Choudalakis and Bill Ashmanskas provided critical comments on the mechanics of the information exchange protocol.  Carla Casamona and Amit Gulati helpfully informed the design of the Kn-X user interface.  Susan Burkhardt provided valuable legal context.  Erik Lindemann delivered particularly useful feedback on an initial draft of this article.}.

\end{document}